\documentclass[aps,prd,onecolumn,showpacs,showkeys,amsmath,amssymb,nofootinbib]{revtex4}
\usepackage{graphicx}
\usepackage{dcolumn}
\usepackage{amsmath}
\usepackage{bm}
\usepackage{yhmath}
\usepackage{textcomp}

\begin{document}

\def\Tr{{\rm Tr}}
\newcommand{\sla}{\not\!}

\allowdisplaybreaks

\title{The Pseudoscalar Meson and Heavy Vector Meson Scattering Lengths}
\author{Zhan-Wei Liu}
\email{liuzhanwei@pku.edu.cn} \affiliation{Department of Physics
and State Key Laboratory of Nuclear Physics and Technology\\
Peking University, Beijing 100871, China}

\author{Yan-Rui Liu}
\email{yrliu@th.phys.titech.ac.jp} \affiliation{Department of
Physics, H-27, Tokyo Institute of Technology, Meguro, Tokyo
152-8551, Japan}

\author{Xiang Liu}
\email{xiangliu@lzu.edu.cn} \affiliation{School of Physical
Science and Technology, Lanzhou University, Lanzhou 730000, China}

\author{Shi-Lin Zhu}
\email{zhusl@pku.edu.cn}  \affiliation{Department of Physics
and State Key Laboratory of Nuclear Physics and Technology\\
and Center of High Energy Physics, Peking University, Beijing
100871, China }

\begin{abstract}
We have systematically studied the S-wave pseudoscalar meson and heavy vector meson scattering lengths to the
third order with the chiral perturbation theory, which will be helpful to reveal their strong interaction. For
comparison, we have presented the numerical results of the scattering lengths (1) in the framework of the heavy
meson chiral perturbation theory and (2) in the framework of the infrared regularization. The chiral expansion
converges well in some channels.
\end{abstract}



\maketitle

\pagenumbering{arabic}

\section{Introduction}\label{sec1}

In the past eight years we have witnessed the renaissance of the hadron spectroscopy. Many interesting new
hadron states were discovered experimentally, some of which do not fit into the quark model easily. These new
hadron states include (1) the famous $XYZ$ states, which are either charmonium or charmonium-like states above
the open-charm decay threshold; (2) the narrow charm-strange mesons $D_{s0}(2317), D_{s1}(2460)$ etc.; (3) the
charged Upsilon-like $Z_b$ states recently announced by Belle collaboration \cite{belle2011}. A general feature
of many of these new hadrons is that they are very close to the two open-flavor meson threshold. For example,
$X(3872)$ is very close to the $\bar D D^*$, $\rho J/\psi$ and $\omega J/\psi$ threshold. $D_{s0}(2317)$ is very
close to the $DK$ threshold. These charged $Z_b$ states are very close to the $\bar B^{(*)} B^*$ threshold.
Because of their proximity to the two meson threshold, one may wonder whether some of these new hadron states
are good candidates of loosely bound molecular states composed of two mesons? Or will the coupled-channel effect
between the bare $q\bar q$ state in the quark model and the two meson continuum help lower and push the mass of
the bare quark model state close to the threshold? In order to allow the above two mechanisms work, there must
exist attractive interaction between the two mesons.

\par
Generally speaking, it is very difficult to study the hadron interaction starting from the first principle of
strong interaction, i.e., quantum chromodynamics. Most of such investigations are performed on the lattice
numerically. However, we may turn to chiral perturbation theory for help if one of the interacting mesons is a
light pseudoscalar meson. In this case one can derive the scattering amplitude order by order rigourously. From
the scattering amplitude, we can extract the scattering length, which is directly related to the hadron
interaction. In this work we will study the pseudoscalar meson $\phi$ and heavy vector meson $D^*$ scattering
lengths in order to learn whether there exits attraction between the pseudoscalar meson $\phi$ and heavy vector
meson $D^*$. Such a study will provide valuable information on their interaction to the $D^*K$ system.

\par
Up to now, a few new charmed mesons and their properties have
attracted much interest over the past few years, especially the
extremely narrow $D_{s0}(2317), D_{s1}(2460)$ states. The
$D$($D_s$) and $D^*$($D_s^*$) mesons constitute the lightest
charmed doublet according to heavy quark symmetry. The recently
observed new charmed particle $D_{s0}(2317)$ is speculated to be a
candidate of possible molecular states composed of the $D$ meson
and $K$ meson. The $D K$ interaction is very important for us to
understand the underlying structure of the $D_{s0}(2317)$ meson
\cite{Beveren2003,Dai2008}. There have been some lattice
investigations on the $\pi D$ scattering
\cite{Flynn2007,Aoki2010,Torok2010a}, $K D$ scattering and
$D_{s0}(2317)$ \cite{Gong2011,Dong2009,Bernard2001}. If there are
strongly attractive interactions between them, the $D^*$ meson and
pseudoscalar meson might also form possible molecular states. Our
present study of the $\phi D^*$ scattering with chiral
perturbation theory would be helpful to the future lattice
simulation of the $\phi D^*$ scattering numerically.

\par
Chiral perturbation theory and lattice QCD are widely used to study the hadron interaction in the
nonperturbative region of QCD
\cite{Lee1994,Mojzis1998,Kaiser2001,Cohen2006,Meng2009,Gamermann2007,Torok2010,Scherer2003}. The interaction
between the $D$ meson and the light pseudoscalar meson was studied recently with chiral perturbation theory
\cite{Liu2009,Guo2009,Geng2010}. It is interesting to extend the same formalism to study the interaction of the
heavy vector meson $D^*$ and the light pseudoscalar meson. In this paper, we will calculate the S-wave
scattering lengths of the pseudoscalar meson and $D^*$ meson with the heavy meson chiral perturbation theory
(HM$\chi$PT) and the infrared regularization
(IR)\cite{Yan1992,Cho1993,Wise1992,Casalbuoni1997,Becher1999,Becher2001,Kubis2001,Zhu2001b}. The scattering
length $a_{\phi D^*}$ reflects their interaction. In our formalism $a_{\phi D^*}$ is related to the threshold
$T$-matrix $T_{\phi D^*}$: $T_{\phi D^*}=8\pi (1+\frac{m_\phi}{M_{D^*}}) a_{\phi D^*}$, where $m_\phi$ and
$M_{D^*}$ are the masses of the light pseudoscalar meson and $D^*$ meson respectively.

\par
With the explicit power counting scheme, HM$\chi$PT is a useful tool to investigate the heavy meson interactions
\cite{Yan1992,Cho1993,Wise1992,Casalbuoni1997}. We will expand our calculation by $\epsilon=p/\Lambda_\chi$,
where $p$ represents the momentum of the light pseudoscalar meson, the small residue momentum of the heavy meson
in the nonrelativistic limit, or the mass difference $\delta$ between $D$ and $D^*$ mesons, and $\Lambda_\chi$
represents either the chiral symmetry breaking scale around $4\pi f_\pi$ or the heavy mesons' masses $\ring M$
(about 1900 MeV) in the chiral and heavy quark symmetry limit. The IR scheme is also a useful tool based on
chiral perturbation theory \cite{Becher1999,Becher2001,Kubis2001,Zhu2001b}, which ensures both good power
counting and correct analyticity. IR and HM$\chi$PT generally lead to the same results except that the IR
formalism includes the higher-order infrared parts of the loop graphs \cite{Becher1999}.

\par
This paper is organized as follows. In Sec. II we list the basic
notations, the relevant Lagrangians, and the chiral corrections to
the threshold $T$-matrices with HM$\chi$PT. We present the IR
expressions in Sec. III. The low-energy constants (LECs) are
estimated in Sec. IV. Finally, we give the numerical results and
discussions in Sec. V.

\section{The $T$-matrices with the Heavy Meson Chiral Perturbation Theory}\label{sec2}
We list the Lagrangian of HM$\chi$PT at the leading order here,
\begin{eqnarray}
  \mathcal L^{(2)}_{\phi\phi}&=&f^2 \Tr\left(u_\mu u^\mu +\frac{\chi_+}{4}\right), \\
  \mathcal L^{(1)}_{H\phi}&=&-\langle (i v\cdot \partial H)\bar H \rangle
                         +\langle H v\cdot \Gamma \bar H \rangle
                         +g\langle H \sla u \gamma_5 \bar H\rangle
                         -\frac18 \delta \langle H \sigma^{\mu\nu} \bar H \sigma_{\mu\nu} \rangle,\label{L1}
\end{eqnarray}
where $f$ is the decay constant of the pseudoscalar meson in the
chiral limit, and
\begin{eqnarray}
 && H=\frac{1+\sla v}{2}\left(P^*_\mu\gamma^\mu+iP\gamma_5\right),\quad
 \bar H=\gamma^0 H^\dag \gamma^0 = \left(P^{*\dag}_\mu \gamma^\mu+iP^\dag \gamma_5\right) \frac{1+\sla v}{2},\\
 && P=(D^0, D^+, D_s^+), \quad P^*_\mu=(D^{*0}, D^{*+}, D_s^{*+})_\mu. \label{PPs}
\end{eqnarray}
Heavy quark symmetry is exact only when the heavy quark mass is infinite. In this work we will also
systematically include effects of the explicitly broken heavy quark symmetry through the last term containing
the $D^\ast$ and $D$ mass difference $\delta$ in Eq. (\ref{L1}). The notations read
\begin{equation}
\Gamma_\mu = {i\over 2} [\xi^\dagger, \partial_\mu\xi],\quad
u_\mu={i\over 2} \{\xi^\dagger, \partial_\mu \xi\},\quad \xi =
\exp(i \phi/2f), \quad \chi_\pm =
\xi^\dagger\chi\xi^\dagger\pm\xi\chi\xi,\quad
\chi=\mathrm{diag}(m_\pi^2,\, m_\pi^2,\, 2m_K^2-m_\pi^2),
\end{equation}
\begin{equation}
 \phi=\sqrt2\left(
\begin{array}{ccc}
\frac{\pi^0}{\sqrt2}+\frac{\eta}{\sqrt6}&\pi^+&K^+\\
\pi^-&-\frac{\pi^0}{\sqrt2}+\frac{\eta}{\sqrt6}&K^0\\
K^-&\overline{K}^0&-\frac{2}{\sqrt6}\eta
\end{array}\right).
\end{equation}

\par
The following Lagrangians at the second and third order are used in the calculation of the threshold
$T$-matrices, \footnote{The sign in front of $c_1$ in Eq. (9) of Ref. \cite{Liu2009} should be +. The signs in
Eqs. (12) and (13) should be consequently changed.}
\begin{eqnarray}
  \mathcal L^{(2)}_{H\phi}&=& c_0 \langle H \bar H\rangle \Tr(\chi_+)
                              +c_1 \langle H \chi_+ \bar H\rangle
                              -c_2 \langle H \bar H\rangle \Tr(v\cdot u~v\cdot u)
                              -c_3 \langle H v\cdot u~v\cdot u \bar H\rangle ,\label{L2}\\
  \mathcal L^{(3)}_{H\phi}&=& \kappa_0  \delta \langle H   \bar H\rangle \Tr(\chi_+)
                              +\kappa_1  \delta \langle H \chi_+  \bar H\rangle
                              -\kappa_2  \delta \langle H   \bar H\rangle \Tr(v\cdot u~v\cdot u)
                              -\kappa_3  \delta \langle H v\cdot u~v\cdot u  \bar H\rangle
                            +\kappa \langle H [\chi_-,v\cdot u] \bar H \rangle .\label{L3}
\end{eqnarray}
The $O(\epsilon^2)$ and $O(\epsilon^3)$ Lagrangians could also
contain terms like $\langle H \sigma^{\mu\nu} \bar
H\sigma_{\mu\nu}\rangle \Tr(\chi_+)$, $\langle H\sigma^{\mu\nu}
\chi_+ \bar H\sigma_{\mu\nu}\rangle$. These terms break the heavy
quark symmetry hence are suppressed. They lead to different LECs
$c_i$'s and $\kappa_i$'s for the $\phi D$ and $\phi D^*$ scattering,
although they do not result in the new independent vertices we need.

\par
There are eleven independent $T$-matrices in the pseudoscalar
meson and $D^*$ meson scattering due to the isospin symmetry. The
threshold $T$-matrices start at $O(\epsilon)$, which can be
derived from Eq. (\ref{L1})
\begin{eqnarray}
&&T^{(3/2)}_{\pi D^*}=-\frac{m_\pi}{f_\pi^2}, \quad
T^{(1/2)}_{\pi D^*}=\frac{2m_\pi}{f_\pi^2} , \quad
T^{(1)}_{\pi D_s^*}=0 , \quad
T^{(0)}_{KD^*}=\frac{2m_K}{f_K^2}, \quad
T^{(1)}_{KD^*}=0 , \quad
T^{(1/2)}_{KD_s^*}=-\frac{m_K}{f_K^2} , \nonumber\\
&&T^{(1)}_{\bar K D^*}=-\frac{m_K}{f_K^2}, \quad
T^{(0)}_{\bar K D^*}=\frac{m_K}{f_K^2} , \quad
T^{(1/2)}_{\bar K D_s^*}=\frac{m_K}{f_K^2} , \quad
T^{(1/2)}_{\eta D^*}=0 , \quad
T^{(0)}_{\eta D_s^*}=0,
\end{eqnarray}
where the superscript in the bracket represents the total isospin
of the channel. We express $T$-matrices with the renormalized
decay constants $f_\pi$, $f_K$ and $f_\eta$
\cite{Gasser1985,Liu2011} rather than $f$ here. The difference
could be accounted for through $T$-matrices at $O(\epsilon^3)$ or
higher order.

\par
Similarly we get the results at $O(\epsilon^2)$, \footnote{The constant $C_1$ in $T_{DK}^{(0)}$ of Ref.
\cite{Liu2009} should be $\frac12(3C_1-C_0)$. The corrected $T$-matrix is the same as $T_{KD^*}^{(0)}$ here.}
\begin{eqnarray}
&&T^{(3/2)}_{\pi D^*}=\frac{8  c_0 m_\pi^2+4 c_1 m_\pi^2+2 c_2 m_\pi^2+c_3 m_\pi^2}{f_\pi^2}, \quad
T^{(1/2)}_{\pi D^*}=\frac{8  c_0 m_\pi^2+4 c_1 m_\pi^2+2 c_2 m_\pi^2+c_3 m_\pi^2}{f_\pi^2}, \nonumber\\&&
T^{(1)}_{\pi D_s^*}=\frac{ 8  c_0 m_\pi^2+2c_2 m_\pi^2}{f_\pi^2}, \quad
T^{(0)}_{KD^*}=\frac{ 8  c_0 m_K^2+8 c_1 m_K^2+2c_2 m_K^2+2c_3 m_K^2}{f_K^2},\quad
T^{(1)}_{KD^*}=\frac{8  c_0 m_K^2+2c_2 m_K^2}{f_K^2},\nonumber\\&&
T^{(1/2)}_{KD_s^*}=\frac{8  c_0 m_K^2+4 c_1 m_K^2+2 c_2 m_K^2+c_3 m_K^2}{f_K^2}, \quad
T^{(1)}_{\bar K D^*}=\frac{8  c_0 m_K^2+4 c_1 m_K^2+2 c_2 m_K^2+c_3 m_K^2}{f_K^2},\nonumber\\&&
T^{(0)}_{\bar K D^*}=\frac{8  c_0 m_K^2-4 c_1 m_K^2+2 c_2 m_K^2-c_3 m_K^2}{f_K^2}, \quad
T^{(1/2)}_{\bar K D_s^*}=\frac{8  c_0 m_K^2+4 c_1 m_K^2+2 c_2 m_K^2+c_3 m_K^2}{f_K^2}, \nonumber\\&&
T^{(1/2)}_{\eta D^*}=\frac{24  c_0 m_\eta^2+4 c_1 m_\pi^2+6 c_2 m_\eta^2+c_3 m_\eta^2}{3 f_\eta^2},\quad
T^{(0)}_{\eta D_s^*}=\frac{ 24  c_0 m_\eta^2+32 c_1 m_K^2-16 c_1 m_\pi^2+6 c_2 m_\eta^2
                          +4 c_3 m_\eta^2}{3 f_\eta^2}. \label{T2}
\end{eqnarray}
Here we have used the Gell-Mann-Okubo mass relation
$m_\eta^2=(4m_K^2-m_\pi^2)/3$, which makes the expressions more
concise.

\par
\begin{figure}[!htbp]
\centering
\scalebox{0.75}{\includegraphics{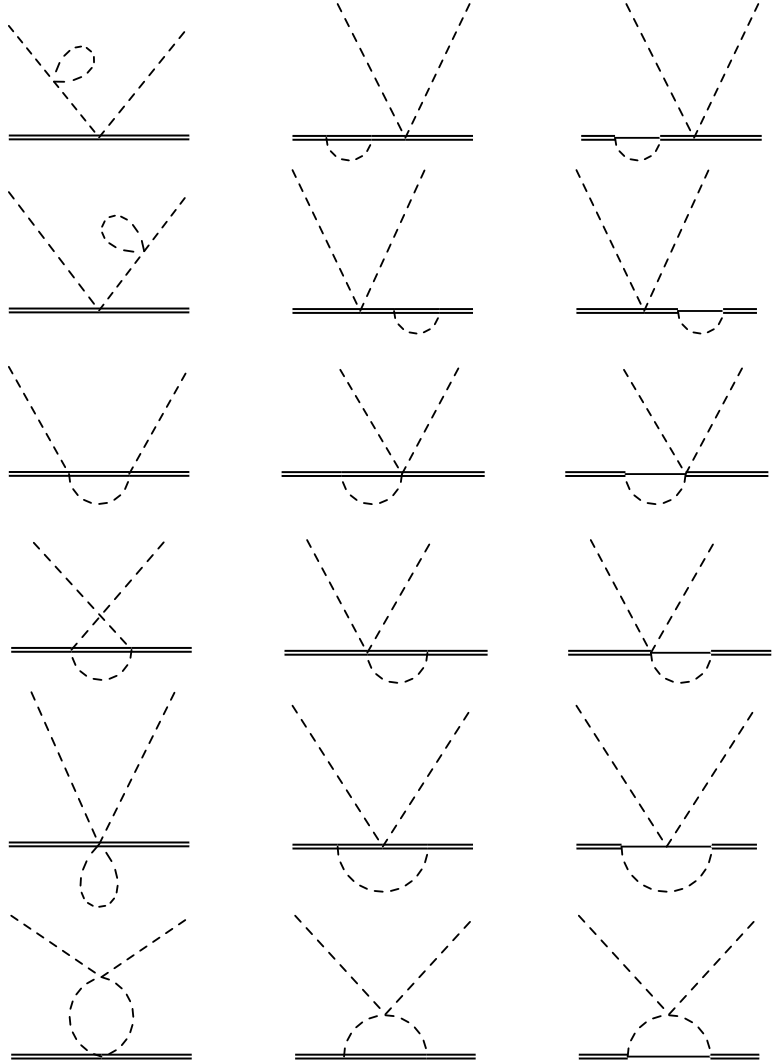}}\\
\caption{Nonvanishing loop diagrams for the pseudoscalar meson and
$D^*$ meson scattering lengths to $O(\epsilon^3)$ with HM$\chi$PT
and IR. The dashed lines, thin solid lines and thick solid lines
represent the pseudoscalar Goldstone bosons, $D$ mesons  and $D^*$
mesons, respectively.} \label{LoopDiag}
 \end{figure}

The $T$-matrices contain contributions from both the tree and loop diagrams. We show all eighteen loop diagrams
which contribute to the threshold $T$-matrix at $O(\epsilon^3)$ in Fig. \ref{LoopDiag}. We calculate them with
the dimensional regularization and modified minimal subtraction. More specifically, for the unstable $D^*$ meson
we renormalize its wave function at the point $\bar r$
\begin{equation}
  Z_{D^*}=1+\left.\frac{d\Pi_{D^*}( r)}{d( r)}\right|_{ r=\bar r},
\end{equation}
where $\Pi_{D^*}( r)$ is the one-particle irreducible $D^*$ self-energy, $r$ is the remaining momentum $r\equiv
v \cdot p-\ring M$, and $\bar r$ is the complex pole of the propagator $ \bar r-\delta-\Pi_{D^*}(\bar r)=0$. The
divergences from loops can be absorbed after the wave function renormalization and redefinitions of $\kappa_i$,
\begin{equation}
  4\kappa_0+\kappa_2= \frac{2g^2 L}{9f^2}+4\kappa_0^r+\kappa_2^r,\quad
  \kappa_1=\frac{5g^2 L}{12f^2} +\kappa_1^r,\quad
  \kappa_3=- \frac{3g^2 L}{f^2} +\kappa_3^r,\quad
  \kappa=\frac{ 3L}{4f^2} +\kappa^r,
\end{equation}
where
\begin{equation}
  L=\frac{\lambda^{D-4}}{16\pi^2}\left\{\frac1{D-4}+\frac12(\gamma_E-1-\ln 4\pi)\right\}.\quad
     (\text{Euler constant~}\gamma_E=0.5772157)
\end{equation}
Here $\lambda$ is the scale of the dimensional regularization. We
will set it at $4\pi f_\pi$, $4\pi f_K$ and $4\pi f_\eta$
respectively for the pion-, kaon- and $\eta$-scattering.

\par
In order to make the expressions short, we introduce the following
notations and functions:
\begin{eqnarray}
  J&=&-\frac{g^2}{24 \pi ^2 \left({m_\eta}^2-{m_\pi}^2\right)f^4}
    \left(2\pi  {m_\eta}^3+2 ({m_\eta}^2-\delta ^2)^{3/2} \cos ^{-1}(-\frac{\delta }{{m_\eta}})
  +3 {m_\eta}^2 \delta  \log \frac{{m_\eta}}{\lambda }-2 {m_\eta}^2 \delta
  \right.\nonumber\\
  &&\qquad
  -2 \delta ^3 \log \frac{{m_\eta}}{\lambda }
  -2\pi  {m_\pi}^3-2 (\delta ^2-{m_\pi}^2)^{3/2} \log \frac{{m_\pi}}{\lambda }
  +2 (\delta ^2-{m_\pi}^2)^{3/2} \log \frac{\sqrt{\delta ^2-{m_\pi}^2}+\delta }{\lambda }
  \nonumber\\
  &&\qquad\left.
  -2 i \pi  \delta ^2 \sqrt{\delta ^2-{m_\pi}^2}
  +2 i \pi  {m_\pi}^2 \sqrt{\delta ^2-{m_\pi}^2}
  -3 {m_\pi}^2 \delta  \log \frac{{m_\pi}}{\lambda }+2 {m_\pi}^2 \delta
  +2 \delta ^3 \log \frac{{m_\pi}}{\lambda }\right)  ,\nonumber\\
 W(m)&=&
 -\frac{g^2}{16 \pi ^2 f^4}
 \begin{cases}
 2 \sqrt{m^2-\delta ^2} \cos ^{-1}\left(-\frac{\delta }{m}\right)
        +2 \delta  \log \left(\frac{m}{\lambda }\right)+2\pi  m-\delta
                                                                     & m>\delta  \\
 2\delta\log\frac m \lambda
 +2\sqrt{\delta^2-m^2}\left(\log\frac{\sqrt{\delta^2-m^2}+\delta}{m} -i \pi\right) +2\pi m-\delta
                             & m\leq \delta
 \end{cases}, \nonumber\\
 V(m,\omega)&=&
  \frac{\omega^3 \log \frac{m}{\lambda }}{\pi ^2 f^4}-\frac{\omega^3}{2 \pi ^2 f^4}
  -\frac{ \omega^2}{\pi^2 f^4}
    \begin{cases}
      -\sqrt{m^2-\omega^2} \cos ^{-1}\left(-\frac{\omega}{m} \right)& m^2\geq \omega^2 \\
      \sqrt{\omega^2-m^2} \log \frac{\sqrt{\omega^2-m^2}-\omega}{m} & m^2<\omega^2, \omega<0 \\
      \sqrt{\omega^2-m^2} \left(-\log \frac{\sqrt{\omega^2-m^2}+\omega}{m}+i \pi \right)& m^2<\omega^2, \omega\geq 0
   \end{cases}.
  \end{eqnarray}

In the third-order $T$-matrices there are some terms proportional
to those in Eq. (\ref{T2}). Therefore, we divide the $T$-matrices
into two parts,
\begin{equation}
  T=\tilde T+T_2,
\end{equation}
where $T_2$ can be obtained from Eq. (\ref{T2}) through the
following replacement:
\begin{equation}
  c_0 \rightarrow \kappa_0^r \delta, \quad
  c_1 \rightarrow \kappa_1^r \delta, \quad
  c_2 \rightarrow \kappa_2^r \delta, \quad
  c_3 \rightarrow \kappa_3^r \delta. \quad
\end{equation}
The remaining $\tilde T$ at $O(\epsilon^3)$ reads
\begin{eqnarray}
\tilde T^{(3/2)}_{\pi D^*}&=&-\frac{1}{8}
V(m_K,-m_\pi)-\frac{3}{8} V(m_\pi,-m_\pi)
          -\frac{1}{8}V(m_\pi,m_\pi)+\frac{1}{18} m_\pi^2 W(m_\eta)-\frac{1}{2} m_\pi^2 W(m_\pi)+2 V_1
          -\frac{8 m_\pi^3\kappa^r}{f_\pi^2}, \nonumber\\
\tilde T^{(1/2)}_{\pi D^*}&=&\frac{1}{16}
V(m_K,-m_\pi)-\frac{3}{16} V(m_K,m_\pi)-\frac{1}{2} V(m_\pi,m_\pi)
          +\frac{1}{18}m_\pi^2 W(m_\eta)-\frac{1}{2} m_\pi^2 W(m_\pi)-4 V_1
          +\frac{16 m_\pi^3\kappa^r}{f_\pi^2}, \nonumber\\
\tilde T^{(1)}_{\pi D_s^*}&=&-\frac{1}{8} V(m_K,-m_\pi)-\frac{1}{8} V(m_K,m_\pi)+\frac{2}{9} m_\pi^2 W(m_\eta),  \nonumber\\
\tilde T^{(0)}_{KD^*}&=&-\frac{3}{8} V(m_\eta,m_K)-\frac{1}{8}
V(m_K,-m_K)-\frac{1}{2} V(m_K,m_K)-J m_K^2
         -\frac{1}{9} m_K^2 W(m_\eta)-3 V_2-3 V_3+V_4
         +\frac{16m_K^3\kappa^r}{f_K^2},\nonumber\\
\tilde T^{(1)}_{KD^*}&=&-\frac{1}{8} V(m_\pi,m_K)-\frac{1}{8}
V(m_K,-m_K)+\frac{J m_K^2}{3}
         -\frac{1}{9} m_K^2 W(m_\eta)-V_2+V_3+\frac{V_4}{3}, \nonumber\\
\tilde T^{(1/2)}_{KD_s^*}&=&-\frac{3}{16}
V(m_\eta,-m_K)-\frac{3}{16} V(m_\pi,-m_K)-\frac{1}{8} V(m_K,-m_K)
           -\frac{1}{8}  V(m_K,m_K)-\frac{4}{9} m_K^2 W(m_\eta)+3 V_2
           -\frac{8 m_K^3\kappa^r}{f_K^2} ,  \nonumber\\
\tilde T^{(1)}_{\bar K D^*}&=&-\frac{3}{16}
V(m_\eta,-m_K)-\frac{1}{16} V(m_\pi,-m_K)-\frac{1}{4} V(m_K,-m_K)
           -\frac{1}{8}V(m_K,m_K)-\frac{J m_K^2}{3}-\frac{1}{9} m_K^2 W(m_\eta)\nonumber\\
           &&+2 V_2+V_3 -\frac{2 V_4}{3}
           -\frac{8m_K^3\kappa^r}{f_K^2},   \nonumber\\
\tilde T^{(0)}_{\bar K D^*}&=&\frac{3}{16}
V(m_\eta,-m_K)-\frac{3}{16} V(m_\pi,-m_K)+\frac{1}{4} V(m_K,-m_K)
            -\frac{1}{8}V(m_K,m_K)+J m_K^2-\frac{1}{9} m_K^2 W(m_\eta)-3 V_3\nonumber\\&&
            +\frac{8m_K^3\kappa^r}{f_K^2},    \nonumber\\
\tilde T^{(1/2)}_{\bar K D_s^*}&=&-\frac{3}{16}
V(m_\eta,m_K)-\frac{3}{16} V(m_\pi,m_K)
             -\frac{1}{8}V(m_K,-m_K)-\frac{1}{8} V(m_K,m_K)-\frac{4}{9} m_K^2 W(m_\eta)-3 V_2
             +\frac{8m_K^3\kappa^r}{f_K^2},  \nonumber\\
\tilde T^{(1/2)}_{\eta D^*}&=&-\frac{3}{16}
V(m_K,-m_\eta)-\frac{3}{16} V(m_K,m_\eta)-\frac{2}{3} m_K^2 W(m_K)
             +\frac{2}{9}m_\eta^2 W(m_\eta)-\frac{1}{18} m_\pi^2 W(m_\eta)+\frac{1}{2} m_\pi^2 W(m_\pi),\nonumber\\
\tilde T^{(0)}_{\eta D_s^*}&=&-\frac{3}{8}
V(m_K,-m_\eta)-\frac{3}{8} V(m_K,m_\eta)-\frac{4}{3} m_K^2 W(m_K)
             +\frac{8}{9}m_\eta^2 W(m_\eta)-\frac{2}{9} m_\pi^2 W(m_\eta),  \label{Tloop}
\end{eqnarray}
where $V_1=V_2=V_3=V_4=0$ with HM$\chi$PT but nonzero with IR.

\par

The $T$-matrices of the pseudoscalar meson $\phi$ and $D$ meson scattering with HM$\chi$PT were derived in Ref.
\cite{Liu2009}. For the $\phi D$ case, there is no $\phi D D$-vertex in the leading order. Therefore in the loop
calculation there are no similar diagrams of the third column in Fig. \ref{LoopDiag} where $\phi D$ are the
intermediate states. If we let $\delta \to 0$ and neglect the explicit breaking of heavy quark symmetry, the
$T$-matrices of the pseudoscalar meson and $D^*$ scattering will be the same as those of the pseudoscalar meson
and $D$ meson scattering to $O(\epsilon^3)$ at the threshold \cite{Liu2009}, which is required by the heavy
quark symmetry.

\section{The $T$-matrices with the Infrared Regularization Method} \label{sec3}

For the IR scheme, we use the heavy meson Lagrangian with the
relativistic form at the leading order,
\begin{eqnarray}
  \mathcal L^{(1)}_{H\phi}&=&  \mathcal D_\mu \tilde P \mathcal D^\mu \tilde P^\dag
                       -\ring M^2  \tilde P \tilde P^\dag
                      - \mathcal D_\mu \tilde P^{*\nu} \mathcal D^\mu \tilde P^{*\dag}_\nu
                      +(\ring M+\delta)^2  \tilde P^{*\nu} \tilde P^{*\dag}_\nu \nonumber\\
                  &&
                      + i 2 g \ring M ( \tilde P^*_\mu u^\mu \tilde P^\dag-\tilde P u^\mu \tilde P^{*\dag}_\mu)
                      + g (\tilde P^*_\mu u_\alpha \mathcal D_\beta \tilde P^{*\dag}_\nu
                                 -\mathcal D_\beta \tilde P^*_\mu u_\alpha \tilde P^{*\dag}_\nu)
                                 \epsilon^{\mu \nu \alpha \beta}  ,
\end{eqnarray}
where heavy quark symmetry is also assumed to relate the couplings
of the $\pi D^* D$-vertex and $\pi D^* D^*$-vertex, and
\begin{equation}
  \tilde P=\frac P{\sqrt{\ring M}}, \quad
  i\mathcal D_\mu \tilde P_a=i\partial_\mu \tilde P_a-\Gamma^{ba}_\mu \tilde P_b, \quad
  i\mathcal D_\mu \tilde P_a^\dag=i\partial_\mu \tilde P_a^\dag+\Gamma^{ab}_\mu \tilde P_b^\dag. \qquad
  \text{(Similar for $\tilde P^*$.)}
\end{equation}
The second-order and third-order Lagrangians we use are the same as the relevant terms in Eqs. (\ref{L2},
\ref{L3}), but some coefficients should be redefined to fit the experimental data,
\begin{eqnarray}
  \mathcal L^{(2)}_{H\phi}&=&2\ring M( c_0 \tilde P^{*\mu} \tilde P^{*\dag}_\mu \Tr(\chi_+)
                              +c_1 \tilde P^{*\mu} \chi_+ \tilde P^{*\dag}_\mu
                              -c_2 \tilde P^{*\mu} \tilde P^{*\dag}_\mu \Tr(u\cdot u)
                              -c_3 \tilde P^{*\mu} u\cdot u \tilde P^{*\dag}_\mu),\label{IRL2}\\
  \mathcal L^{(3)}_{H\phi}&=& 2\ring M(\kappa_0  \tilde P^{*\mu} \tilde P^{*\dag}_\mu \Tr(\chi_+)  \delta
                              +\kappa_1  \tilde P^{*\mu} \chi_+  \tilde P^{*\dag}_\mu  \delta
                              -\kappa_2  \tilde P^{*\mu} \tilde P^{*\dag}_\mu \Tr( u\cdot u) \delta
                              -\kappa_3  \tilde P^{*\mu}  u\cdot u  \tilde P^{*\dag}_\mu \delta ) \nonumber\\&&
                            +i\ring M\kappa \left(\mathcal D_\nu \tilde P^{*\mu} [\chi_-,u^\nu] \tilde P^{*\dag}_\mu
                                          -\tilde P^{*\mu} [\chi_-,u^\nu] \mathcal D_\nu \tilde P^{*\dag}_\mu \right).\label{IRL3}
\end{eqnarray}

\par
The $T$-matrices are nearly the same as those of HM$\chi$PT except that the expressions of $J$, $W(m)$,
$V(m,\omega)$, $V_1$, $V_2$, $V_3$, and $V_4$ are more complicated.  We list their definitions in the infrared
scheme in Eq. (\ref{IRfunc}) in the Appendix. We have also verified that the results with IR are the same as
those with HM$\chi$PT when $\ring M$ approaches to infinity.

\section{Low-Energy Constants}\label{sec4}

The difference of our results between HM$\chi$PT and IR originates
at the third order due to different loop integrals. We use the
same LECs at the first and second order for both HM$\chi$PT and
IR. At the leading order, we have \cite{PDG2010}
\begin{eqnarray}
 &&m_\pi=139~{\rm MeV}, \quad
    m_K=494~{\rm MeV},  \quad
    \delta=142~{\rm MeV},\nonumber \\
 &&f_\pi=92~{\rm MeV},\quad
   f_K=113~{\rm MeV},\quad
   f_\eta=1.2 f_K, \quad
   g=0.59. \nonumber
\end{eqnarray}
At the second order, from the mass splitting between heavy mesons we get\footnote{The coefficient 4 in Eq. (33)
of Ref. \cite{Liu2009} should be 8. The correct values of $c_1$ are 0.12 GeV$^{-1}$ in Eq. (\ref{c1}) and 0.10
GeV$^{-1}$ in Eq. (\ref{bci}) of this paper.}
\begin{equation}
  c_1=\frac1{16} \frac{M^2_{D_s}-M^2_D+M^2_{D^*_s}-M^2_{D^*}}{\ring M (m^2_K-m^2_\pi)}=0.12~ {\rm GeV^{-1} }.
  \label{c1}
\end{equation}
In order to obtain other LECs that can not be determined from the available experimental data, we resort to the
resonance saturation model \cite{Ecker1989,Bernard1993a}.

At $O(\epsilon^2)$ only the light unflavored mesons with $J^P=0^+$
and charmed mesons with $J^P=1^+$ contribute to the $\phi\phi D^*
D^*$ vertex at threshold. Thus we consider the scalar singlet
$\sigma$$(\sigma(600))$, the scalar octet $\kappa$($\kappa(800),
a_0(980), f_0(980)$), and $D_{s1}(2460)$ vector triplet in this
section. In the Appendix \ref{appendixA} we will discuss the
uncertainty of the LECs at this order.

\par
Here we list the corresponding effective Lagrangians,
\begin{eqnarray}
  {\cal L}_{\sigma \pi\pi}&=&4\tilde c_d \Tr(u\cdot u) \sigma+\tilde c_m \Tr(\chi_+) \sigma , \nonumber\\
  {\cal L}_{\sigma P^* P^*} &=& c_{\sigma} P^{*\mu}P^{*\dag}_\mu \sigma.
\end{eqnarray}
Integrating the $\sigma$ meson out through the t-channel we get
\begin{equation}
  {\cal L}_{\text{eff}}^{\sigma}\sim
                      \frac{2c_{\sigma} \tilde c_m}{m_{\sigma}^2} \Tr(\chi_+) P^{*\mu}P^{*\dag}_\mu
                      +\frac{8c_{\sigma} \tilde c_d}{m_{\sigma}^2} \Tr(v\cdot u~ v\cdot u) P^{*\mu} P^{*\dag}_\mu
                      \label{Lsigma}
\end{equation}
Similarly, from the Lagrangians of the scalar octet $\kappa$
\begin{eqnarray}
  {\cal L}_{\kappa \pi\pi}&=&4c_d \Tr(u\cdot u \kappa)+c_m \Tr(\chi_+ \kappa), \nonumber\\
  {\cal L}_{\kappa P^* P^*} &=& c_{\kappa} P^{*\mu} \kappa P^{*\dag}_\mu,
\end{eqnarray}
one obtains
\begin{equation}
  {\cal L}_{\text{eff}}^{\kappa}\sim
                      - \frac{2c_{\kappa} c_m}{3m_{\kappa}^2} \Tr(\chi_+) P^{*\mu} P^{*\dag}_\mu
                      +\frac{2c_{\kappa} c_m}{m_{\kappa}^2} P^{*\mu}  \chi_+ P^{*\dag}_\mu
                      - \frac{8c_{\kappa} c_d}{3m_{\kappa}^2} \Tr(v\cdot u~ v\cdot u) P^{*\mu} P^{*\dag}_\mu
                      +\frac{8c_{\kappa} c_d}{m_{\kappa}^2}
                                      P^{*\mu} v\cdot u~ v\cdot u P^{*\dag}_\mu
                      \label{Lkappa}
\end{equation}

Integrating $D_{s1}(2460)$ out of the following interacting
Lagrangian,
\begin{eqnarray}
  \mathcal L_{D_{s1}(2460)}&=&G_1 \left( D_{s1}^{\mu\dag}(2460) u_\nu i \partial^\nu \tilde P^*_\mu
                             -i \partial^\nu\tilde P_\mu^{*\dag} u_\nu D_{s1}^{\mu}(2460) \right)
                  +G_2 \left(-i \partial_\nu D_{s1}^{\mu\dag}(2460) u_\nu \tilde P^*_\mu
                             +\tilde P^*_\mu u_\nu i \partial_\nu D_{s1}^{\mu}(2460) \right)\nonumber\\
                &=&  (G_1+G_2)\left( D_{s1}^{\mu\dag}(2460) u_\nu i \partial^\nu \tilde P^*_\mu
                             -i \partial^\nu\tilde P_\mu^{*\dag} u_\nu D_{s1}^{\mu}(2460) \right)
                    +O(\epsilon^2),\label{Lvec}
\end{eqnarray}
one gets
\begin{equation}
  \mathcal L_{\text{eff}}^{D_{s1}(2460)}=-\frac{|G_1+G_2|^2 M_{D^*}}{M_{D_{s1}(2460)}^2-M_{D^*}^2}
                           P^{*\mu} v\cdot u~ v\cdot u P^{*\dag}_\mu.  \label{LDs12460}
\end{equation}
The effective coupling constants $|G_1+G_2|$ were estimated with
QCD sum rule approach in Ref. \cite{Colangelo1998}:
$|G_1+G_2|=1.2\pm0.2$.

\par
Adding the above effective Lagrangians (\ref{Lsigma}),
(\ref{Lkappa}) and (\ref{LDs12460}) together, one can estimate the LECs by comparing
the sum with the relevant terms in Eq. (\ref{L2}),
\begin{equation}
  c_0=\frac{c_{\sigma} \tilde c_m}{m_{\sigma}^2}- \frac{c_{\kappa} c_m}{3m_{\kappa}^2}, \quad
  c_1=\frac{c_{\kappa} c_m}{m_{\kappa}^2},  \quad
  c_2=-\frac{4c_{\sigma} \tilde c_d}{m_{\sigma}^2}+ \frac{4c_{\kappa} c_d}{3m_{\kappa}^2} , \quad
  c_3=-\frac{4c_{\kappa} c_d}{m_{\kappa}^2}+\frac{|G_1+G_2|^2 M_{D^*}}{2(M_{D_{s1}(2460)}^2-M_{D^*}^2)}. \label{cValue}
\end{equation}

\par
For the broad resonances $\sigma(600)$ and $\kappa(800)$, we use
the masses and widths extracted from a model-independent way
\cite{Caprini2006,Descotes-Genon2006},
\begin{equation}
  m_\sigma=441^{+16}_{-8}~{\rm MeV}, \quad
  \Gamma_\sigma=544^{+18}_{-25}~{\rm MeV}; \quad
  m_{\kappa(800)}=658\pm 13~{\rm MeV}, \quad
  \Gamma_{\kappa(800)}=557\pm 24~{\rm MeV}.
\end{equation}
In our numerical analysis, we take $m_\kappa=820~{\rm MeV}$ for
illustration.

\par
For the coupling constants $c_d$ and $c_m$, we use
\cite{Ecker1989}
\begin{equation}
  \left|c_d\right|=3.2\times10^{-2}~{\rm GeV}, \quad
  \left|c_m\right|=4.2\times10^{-2}~{\rm GeV}, \quad
  c_d c_m>0.
\end{equation}
Although there is no empirical value of $c_\kappa$, we may get it
by comparing the $c_1$'s obtained in different ways in Eq.
(\ref{c1}) and Eq. (\ref{cValue})
\begin{equation}
  \left|c_\kappa\right|=1.9, \quad c_\kappa c_m > 0.
\end{equation}
Moreover the coupling constants should obey the nonet relations in
the large $N_c$ limit,
\begin{equation}
  \tilde c_d=\frac{\zeta}{\sqrt 3}c_d, \quad
  \tilde c_m=\frac{\zeta}{\sqrt 3}c_m, \quad
  c_\sigma=\frac{\zeta}{\sqrt 3}c_\kappa, \quad
  \zeta=\pm 1. \label{cwithctilde}
\end{equation}
In this way, we get the LECs at $O(\epsilon^2)$,
\begin{equation}
  c_0=0.10~ {\rm GeV^{-1} } ,\quad
  c_1=0.12~ {\rm GeV^{-1} } ,\quad
  c_2=-0.30~ {\rm GeV^{-1} },\quad
  c_3=0.42~ {\rm GeV^{-1} }.\quad \label{cNvalue}
\end{equation}

\par
The resonance saturation method may bring large uncertainty in the
determination of the LECs at the third order. We take the value of
$\kappa^r$ in Ref \cite{Liu2009}
\begin{equation}
  \kappa^r=-0.33~{\rm{GeV^{-2}}},
\end{equation}
which is obtained by fitting the lattice QCD results \cite{Liu2008}. We simply assume the other tree diagram
contributions at $O(\epsilon^3)$ are small and neglect them as done in Refs. \cite{Kaiser2001,Liu2007}.

\section{Numerical results and discussions}\label{sec5}

We show the numerical results of the $T$-matrices order by order and the scattering lengths with HM$\chi$PT in
Table \ref{TMatHM}\footnote{The numerical values in Ref. \cite{Liu2009} also need a few corrections. The
corrections are given here in the form of $\{{\cal O}(p^2)$, Total, Scattering lengths$\}_{T}$. In Table I, they
are $\{-14.2,1.1,0.04\}_{T_{DK}^{(0)}}$, $\{6.4,7.5+5.5i,0.23+0.17i\}_{T_{D\eta}}$, and
$\{-6.7,-6.2+11.1i,-0.19+0.35i\}_{T_{D_s\eta}}$. In Table II, they are
$\{-14.2,0.5,0.02\}_{T_{\bar{B}K}^{(0)}}$, $\{6.4,7.3+5.5i,0.26+0.20i\}_{T_{\bar{B}\eta}}$, and
$\{-6.9,-6.6+11.1i,-0.24+0.35i\}_{T_{\bar{B}_s\eta}}$. In Table III, they are
$\{-15.1,-0.6,-0.02\}_{T_{DK}^{(0)}}$, $\{6.1,7.3+5.5i,0.22+0.17i\}_{T_{D\eta}}$, and
$\{-7.4,-7.0+11.1i,-0.22+0.35i\}_{T_{D_s\eta}}$. To get a positive $a_{D_s\eta}$ and nearly vanishing
$a_{D_s\pi}$, one requires $C_1>0.8$ GeV$^{-1}$ and $C_0<4.2$ GeV$^{-1}$.}. The positive real parts of
$a^{(1/2)}_{\pi D^*}$, $a^{(0)}_{K D^*}$, $a^{(0)}_{\bar K D^*}$, $a^{(1/2)}_{\bar K D^*_s}$, $a^{(1/2)}_{\eta
D^*}$ and $a^{(0)}_{\eta D^*_s}$ indicate that the interactions are attractive for these channels.
From Table \ref{TMatHM}, we see that the chiral expansion of the pion channels converges well.
The loop diagrams contribute
largely to the kaon and eta channels due to the large mass of kaon and eta. But luckily they are cancelled by
the tree diagram at $O(\epsilon^3)$, which makes the whole result convergent.

\begin{table}
\caption{The threshold $T$-matrices for the pseudoscalar meson and
$D^*$ meson scattering order by order in units of fm with
HM$\chi$PT. }\label{TMatHM}
\begin{tabular}{ccc|ccc|cc}
\hline
                             &$O(\epsilon^1)$   &$O(\epsilon^2)$   &\multicolumn{3}{c|}{ $O(\epsilon^3)$ }                                  &Total             &Scattering length                \\
                             &                  &                  &loop                   &tree                   &total                   &                  &                                \\
\hline
$T^{(3/2)}_{\pi D^*}$        & -3.2             & 0.5              & -1.-0.0096$i$         & 0.17                  & -0.88-0.0096$i$        & -3.6-0.0096$i$   & -0.13-0.00036$i$       \\
$T^{(1/2)}_{\pi D^*}$        & 6.5              & 0.5              & 0.53-0.0096$i$        & -0.33                 & 0.19-0.0096$i$         & 7.1-0.0096$i$    & 0.27-0.00036$i$        \\
$T^{(1)}_{\pi D_s^*}$        & 0                & 0.09             & -1.1                  & 0                     & -1.1                   & -1               & -0.039                 \\
\hline
$T^{(0)}_{KD^*}$             & 15               & 7.5              & 11.-0.00016$i$        & -9.8                  & 1.1-0.00016$i$         & 24.-0.00016$i$   & 0.76-$5.2\times10^{-6}i$  \\
$T^{(1)}_{KD^*}$             & 0                & 0.75             & -1.5+5.6$i$           & 0                     & -1.5+5.6$i$            & -0.7+5.6$i$      & -0.022+0.18$i$         \\
$T^{(1/2)}_{KD_s^*}$         & -7.6             & 4.1              & -5.9                  & 4.9                   & -0.98                  & -4.5             & -0.14                  \\
$T^{(1)}_{\bar K D^*}$       & -7.6             & 4.1              & -7.4-0.000054$i$      & 4.9                   & -2.5-0.000054$i$       & -5.9-0.000054$i$ & -0.19-$1.7\times10^{-6}i$ \\
$T^{(0)}_{\bar K D^*}$       & 7.6              & -2.6             & 8.8+0.00016$i$        & -4.9                  & 3.9+0.00016$i$         & 8.9+0.00016$i$   & 0.29+$5.2\times10^{-6}i$  \\
$T^{(1/2)}_{\bar K D_s^*}$   & 7.6              & 4.1              & 4.+8.3$i$             & -4.9                  & -0.86+8.3$i$           & 11.+8.3$i$       & 0.35+0.27$i$           \\
\hline
$T^{(1/2)}_{\eta D^*}$       & 0                & 1.2              & 0.46+3.$i$            & 0                     & 0.46+3.$i$             & 1.7+3.$i$        & 0.051+0.094$i$         \\
$T^{(0)}_{\eta D_s^*}$       & 0                & 5.8              & 0.0036+6.1$i$         & 0                     & 0.0036+6.1$i$          & 5.8+6.1$i$       & 0.18+0.19$i$           \\

\hline
\end{tabular}
\end{table}

\par
We compare the loop contribution between the HM$\chi$PT and IR
scheme in Table \ref{LoopVS}. For both cases the dominant loop
contributions are those with the intermediate state $D^*$ meson.
The numerical results are similar in the pion-scattering channels
with these two different schemes. But the results differ greatly
in the eta scattering channels.

\begin{table}
\caption{Comparison of the $T$-matrices from the loop diagrams for the pseudoscalar meson and $D^*$ meson
scattering between HM$\chi$PT and IR.}\label{LoopVS}
\begin{tabular}{c|cc@{~~\textbrokenbar~~}cc|cc|cc}
\hline
                                  &\multicolumn{4}{c|}{ Intermediate state: $D$ meson}                                                                        &\multicolumn{2}{c|}{ Intermediate state: $D^*$ meson}   &\multicolumn{2}{c}{Loop:      total}                  \\
\hline
                                  &\multicolumn{2}{c@{~~\textbrokenbar~~}}{HM$\chi$PT}        &\multicolumn{2}{c|}{IR}                                        &HM$\chi$PT                    & IR                       & HM$\chi$PT                   & IR                     \\
                                  &$\delta=142$~MeV         &$\delta\rightarrow 0$            &$\delta=142$~MeV         &$\delta\rightarrow0$                 &$\delta=142$~MeV              &$\delta=142$~MeV          &$\delta=142$~MeV              &$\delta=142$~MeV        \\
\hline
$T^{(3/2)}_{\pi D^*}$             &-0.053-0.0096$i$         &0.014                            &-0.043-0.0076$i$         & 0.0045                              &-0.99                         &-0.84                     &-1.-0.0096$i$                 &-0.88-0.0076$i$         \\
$T^{(1/2)}_{\pi D^*}$             &-0.053-0.0096$i$         &0.014                            &-0.05-0.0093$i$          & 0.031                               &0.58                          &0.3                       &0.53-0.0096$i$                &0.25-0.0093$i$          \\
$T^{(1)}_{\pi D_s^*}$             &-0.043                   &-0.046                           &-0.03                    & -0.04                               &-1.1                          &-0.88                     &-1.1                          &-0.91                   \\
\hline
$T^{(0)}_{KD^*}$                  &0.69-0.00016$i$          &0.93                             &0.46-0.00015$i$          & 0.76                                &10.                           &7.5                       &11.-0.00016$i$                &8.-0.00015$i$           \\
$T^{(1)}_{KD^*}$                  &-0.076+0.000054$i$       &-0.14                            &-0.014+0.0019$i$         & -0.18                               &-1.4+5.6$i$                   &-3.6+2.9$i$               &-1.5+5.6$i$                   &-3.6+2.9$i$             \\
$T^{(1/2)}_{KD_s^*}$              &0.46                     &0.51                             &0.29                     & 0.43                                &-6.3                          &-13.                      &-5.9                          &-13.                    \\
$T^{(1)}_{\bar K D^*}$            &0.31-0.000054$i$         &0.4                              &0.17-0.00099$i$          & 0.38                                &-7.7                          &-14.                      &-7.4-0.000054$i$              &-14.-0.00099$i$         \\
$T^{(0)}_{\bar K D^*}$            &-0.46+0.00016$i$         &-0.68                            &-0.33-0.0027$i$          & -0.51                               &9.3                           &11.                       &8.8+0.00016$i$                &11.-0.0027$i$           \\
$T^{(1/2)}_{\bar K D_s^*}$        &0.46                     &0.51                             &0.31                     & 0.42                                &3.6+8.3$i$                    &1.1+4.4$i$                &4.+8.3$i$                     &1.5+4.4$i$              \\
\hline
$T^{(1/2)}_{\eta D^*}$            &0.14+0.0021$i$           &0.16                             &0.088+0.0018$i$          & 0.13                                &0.32+3.$i$                    &-2.9+1.6$i$               &0.46+3.$i$                    &-2.8+1.6$i$             \\
$T^{(0)}_{\eta D_s^*}$            &-0.022                   &0.013                            &-0.015                   & 0.016                               &0.025+6.1$i$                  &-6.2+3.2$i$               &0.0036+6.1$i$                 &-6.2+3.2$i$             \\
\hline
\end{tabular}
\end{table}

The mass difference $\delta$ affects our results only through the
intermediate $D$ meson in the loop diagrams. For comparison, we
list the $\delta$-dependent part of the $T$-matrices in Table
\ref{LoopVS} when $\delta\rightarrow0$ and $\delta=142$~MeV. We
notice that the correction from the heavy quark symmetry breaking
in the loop diagrams with the $D$ meson intermediate state is
small in the $K D^*$ channels. However, such a correction is
significant in the $\pi D^*{}^{I=3/2}$, $\pi D^*{}^{I=1/2}$ and
$\eta D_s^*{}^{I=0}$ channels.

From Table \ref{LoopVS}, the IR scheme lowers the loop
contribution in the channels $T^{(3/2)}_{\pi D^*}$,$T^{(1/2)}_{\pi
D^*}$,$T^{(1)}_{\pi D_s^*}$,$T^{(0)}_{KD^*}$, and $T^{(1/2)}_{\bar
K D_s^*}$. The $T$-matrices of the $\phi D$ scattering in the
nonrelativistic $\chi$PT were compared with those in the
relativistic $\chi$PT with the extended-on-mass-shell
renormalization schemes in Ref. \cite{Geng2010}. The relativistic
effect would also lower the loop contribution in some channels
such as $T^{(3/2)}_{\pi D}$,$T^{(1/2)}_{\pi D}$.

The resonance $D_{s1}(2460)$ couples to the $D^*K $ strongly. Its role is similar to that of $\Delta(1232)$ in
the case of the pion nucleon scattering. $D_{s1}(2460)$ is very close to the $D^* K$ threshold and may be quite
important for the $D^* K$ $T$-matrix. In contrast, the non-strange P-wave axial-vector $D$ meson lies well above
the $D^* \pi$ threshold. Its contribution is less important. In this work we have tried to include some of the
corrections from the P-wave axial-vector $D$ meson through the LEC $c_3$. We expect that the results would be
improved particularly for the $D^* K$ channel if $D_{s1}(2460)$ is included as an explicit degree of freedom as
$\Delta(1232)$ in heavy baryon chiral perturbation theory \cite{Jenkins1991,Hemmert1998,Liu2007a}.

\par
We can also study the pseudoscalar meson and $\bar B^*$ meson
scattering with $P^*_\mu$ and $P$ representing $(B^-, \bar B^0,
\bar B^0_s)$ and $(B^{*-}, \bar B^{*0}, \bar B_s^{*0})$,
respectively, in Eq. (\ref{PPs}). The situation is very similar to
the $D^*$ case. With the following different constants
\begin{eqnarray}
 && c_0=0.08~ {\rm GeV^{-1} } ,\quad
  c_1=0.10~ {\rm GeV^{-1} } ,\quad
  c_2=-0.25~ {\rm GeV^{-1} },\quad
  c_3=0.44~ {\rm GeV^{-1} },\nonumber\\
 && \kappa^r=-0.33~{\rm{GeV^{-2}}},\quad
  \ring M=5323~{\rm{MeV}},\quad
  \delta=46~{\rm{MeV}},\label{bci}
\end{eqnarray}
and $g=0.52$ \cite{Ohki2008}, we can also analyze the interaction
of the pseudoscalar meson and $\bar B^*$ meson. The numerical
results with HM$\chi$PT are shown in Table \ref{TB}. They are only
slightly different from those of the pseudoscalar meson and $D^*$
meson scattering.

\begin{table}
\caption{The threshold $T$-matrices for the pseudoscalar meson and
$\bar B^*$ meson scattering order by order in units of fm with
HM$\chi$PT. }\label{TB}
\begin{tabular}{ccc|ccc|cc}
\hline
                                  &$O(\epsilon^1)$   &$O(\epsilon^2)$   &\multicolumn{3}{c|}{ $O(\epsilon^3)$ }                                  &Total       &Scattering length                \\
                                  &                  &                  &loop                   &tree                   &total                   &            &                                \\
\hline
$T^{(3/2)}_{\pi \bar B^*}$        & -3.2             & 0.44             & -1                    & 0.17                  & -0.83                  & -3.6        & -0.14              \\
$T^{(1/2)}_{\pi\bar B^*}$         & 6.5              & 0.44             & 0.58                  & -0.33                 & 0.24                   & 7.1         & 0.28               \\
$T^{(1)}_{\pi \bar B_s^*}$        & 0                & 0.063            & -1.1                  & 0                     & -1.1                   & -1          & -0.04              \\
\hline
$T^{(0)}_{K\bar B^*}$             & 15               & 6.9              & 11                    & -9.8                  & 0.72                   & 23          & 0.83               \\
$T^{(1)}_{K\bar B^*}$             & 0                & 0.53             & -1.4+5.6$i$           & 0                     & -1.4+5.6$i$            & -0.88+5.6$i$& -0.032+0.2$i$      \\
$T^{(1/2)}_{K\bar B_s^*}$         & -7.6             & 3.7              & -6.2                  & 4.9                   & -1.3                   & -5.2        & -0.19              \\
$T^{(1)}_{\bar K \bar B^*}$       & -7.6             & 3.7              & -7.6                  & 4.9                   & -2.7                   & -6.6        & -0.24              \\
$T^{(0)}_{\bar K \bar B^*}$       & 7.6              & -2.6             & 9.1                   & -4.9                  & 4.2                    & 9.2         & 0.34               \\
$T^{(1/2)}_{\bar K \bar B_s^*}$   & 7.6              & 3.7              & 3.7+8.3$i$            & -4.9                  & -1.2+8.3$i$            & 10.+8.3$i$  & 0.37+0.3$i$        \\
\hline
$T^{(1/2)}_{\eta \bar B^*}$       & 0                & 1                & 0.37+3.$i$            & 0                     & 0.37+3.$i$             & 1.4+3.$i$   & 0.049+0.11$i$      \\
$T^{(0)}_{\eta \bar B_s^*}$       & 0                & 5.2              & 0.022+6.1$i$          & 0                     & 0.022+6.1$i$           & 5.2+6.1$i$  & 0.19+0.22$i$       \\
\hline
\end{tabular}
\end{table}

\par
As mentioned in Ref \cite{Liu2009}, it is easy to get pseudoscalar
meson and heavy antimeson scattering lengths with the C-parity
transformation,
\begin{equation}
  T^{(I)}_{\bar H K}=T^{(I)}_{H \bar K}, \quad
  T^{(I)}_{\bar H \bar K}=T^{(I)}_{H K}, \quad
  T^{(I)}_{\bar H \pi/\eta}=T^{(I)}_{H \pi/\eta}, \quad
\end{equation}
where $I$ is the total isospin and $H$($\bar H$) denotes the heavy
meson(antimeson).

\par
In short, we have investigated the pseudoscalar meson and $D^*$
meson scattering lengths to $O(\epsilon^3)$ with HM$\chi$PT and IR
methods. The chiral expansion in the $\pi D^\ast$ channels
converges well. We hope our present study may be helpful to the
possible extrapolation in the future lattice simulation of the
light meson and heavy meson scattering. Our results may also be
useful to the phenomenological investigation of the possible
molecular states composed of one heavy meson and one light meson.

\section*{ Acknowledgments}\label{sec6}

YRL thanks L.S. Geng for communications about results in Ref. \cite{Liu2009}. This project is supported by the
National Natural Science Foundation of China under Grants No. 11075004, No. 11021092, No. 11035006, No.
11047606, No. 10805048, and the Ministry of Science and Technology of China (No. 2009CB825200), and the Ministry
of Education of China (FANEDD under Grants No. 200924, DPFIHE under Grants No. 20090211120029, NCET under Grants
No. NCET-10-0442, the Fundamental Research Funds for the Central Universities under Grants No. lzujbky-2010-69).
YRL was partially supported by JSPS KAKENHI (21.09027).

\appendix
\section{The uncertainty of LECs with resonance saturation method} \label{appendixA}
\par
Other resonances with the same quantum numbers as $\sigma$,
$\kappa$, and $D_{s1}(2460)$ will also contribute to $c_i$'s.
Generally, the heavier the resonance is, the less its contribution
to the LECs. Here we will check the uncertainty which other
resonances would cause.

From Eq. (\ref{cValue}), we see that $\sigma$ does not contribute
to $c_1$ and $c_3$. For the LECs $c_0$ and $c_2$, the contribution $\mathcal K$
from the $\sigma$ and $\kappa$ scales as
\begin{equation}
  \frac{\left|\mathcal K_\kappa \right|}{\left|\mathcal K_\sigma\right|}
  =\frac{m_\sigma^2}{m_\kappa^2}= 0.3. \label{kSigmaKappa}
\end{equation}
In principle the LECs would absorb the contributions from the
$f_0(1370)$ singlet and the $K_0^*$ octet($K_0^*(1430)$,
$a_0(1450)$, $f_0(1500)$) and other heavier resonances similarly.
Similar to Eq. (\ref{kSigmaKappa}),
\begin{equation}
    \frac{\left|\mathcal K_{f_0(1370)} \right|}{\left|\mathcal K_{K_0^*}\right|}
  =\frac{m_{f_0(1370)}^2}{m_{K_0^*}^2}= 0.9. \label{kf0K0star}
\end{equation}

\par
In the channel $f_0(1500)\rightarrow \eta\eta$ and
$f_0(1500)\rightarrow K \bar K$, the decay momentum is $516$ and
$568$ MeV respectively. We can estimate the magnitude of
$c_{d,K_0^*}$ and $c_{m,K_0^*}$ using the experimental partial
width
\begin{equation}
  \Gamma(f_0(1500)\rightarrow \eta\eta)=109^{+7}_{-7}\times (5.1^{+0.9}_{-0.9})\%~{\rm MeV}, \quad
  \Gamma(f_0(1500)\rightarrow K\bar K)=109^{+7}_{-7}\times (8.6^{+1.0}_{-1.0})\%~{\rm MeV},
\end{equation}
and similar effective Lagrangians as in Eq. (\ref{Lkappa}). One
gets
\begin{equation}
   \begin{cases}
      \left|c_{d,K_0^*}\right|\sim 1.9\times10^{-2}~{\rm GeV}, \quad
      \left|c_{m,K_0^*}\right|\sim 3.0\times10^{-2}~{\rm GeV},  &c_{d,K_0^*}~ c_{m,K_0^*}<0; \qquad \text{or} \\
      \left|c_{d,K_0^*}\right|\sim 1.3\times10^{-2}~{\rm GeV}, \quad
      \left|c_{m,K_0^*}\right|\sim 1.5\times10^{-2}~{\rm GeV},  &c_{d,K_0^*}~ c_{m,K_0^*}>0.
   \end{cases}
\end{equation}
Therefore,
\begin{equation}
  \frac{\left|\mathcal K_{K_0^*}\right|}{\left|\mathcal K_\kappa\right|}
     =\frac{|c_{d/m, K_0^*}|}{|c_{d/m}|} \frac{|c_{K_0^*}|}{|c_\kappa|} \frac{m_\kappa^2}{m_{K_0^*}^2}
     \lesssim \frac{|c_{K_0^*}|}{|c_\kappa|} \frac{m_\kappa^2}{m_{K_0^*}^2}
     =0.3 \frac{|c_{K_0^*}|}{|c_\kappa|}
     \sim 0.3 ,   \label{kKostarKappa}
\end{equation}
In the last step we have assumed that the coupling $c_{K_0^*}$ in
the $D^* D^* K_0^*$ vertex is of the same order as $c_\kappa$. Now
we have $\left|\mathcal K_{K_0^*}\right|\sim 0.3\left|\mathcal
K_\kappa\right|$ and $\left|\mathcal K_{f_0(1370)} \right| \sim
0.1\left|\mathcal K_\sigma \right|$. In other words, the
$c_i$s' correction from the heavier resonances is roughly $30\%$.

\par
In principle, heavy vector resonances $D_1(2420)$($D_{s1}(2536)$)
would also give corrections to $c_3$,
\begin{equation}
  \Delta c_3=\frac{\left|G_1+G_2\right|^2_{D_1(2420)} M_{D^*}}{2(M_{D_1(2420)}^2-M_{D^*}^2)}.
\end{equation}
By fitting $\Gamma(D_{s1}(2536)\rightarrow D^* K)<\Gamma(D_{s1}(2536))<2.3~{\rm MeV}$,
one finds $|G_1+G_2|_{D_1(2420)}<0.16$.
So $|\Delta c_3|<0.03~{\rm GeV^{-1}}$, which is less than 10\% of $c_3$ in Eq. (\ref{cNvalue}).

\par
Moreover, the variation of the mass of the $\kappa$ octet from 658
MeV to 985 MeV would introduce the $20\%$ uncertainty to $c_0$ and
$c_2$ and $60\%$ uncertainty to $c_1$ and $c_3$. In short, the
determination of $c_i$'s in Eq.(\ref{cNvalue}) are not accurate.
But their sign and order of magnitude should be reliable.

\section{Some functions and constants in the infrared scheme}\label{appendixB}

We perform the tensor decomposition of the IR integrals as
follows,
\begin{eqnarray}
 && \frac1{i}\int_I \frac{d^d k}{(2\pi)^d}
      \frac{ \{1, k^\mu, k^\mu k^\nu\} } {(k^2-m^2+i\epsilon) \left[(p-k)^2-M^2+i\epsilon\right]}\nonumber\\
 &=&\left\{I^{(0)}(p^2,m^2,M^2),\quad
            p^\mu I^{(1)}(p^2,m^2,M^2),\quad
            g^{\mu\nu}I^{(2)}(p^2,m^2,M^2)+p^\mu p^\nu I^{(3)}(p^2,m^2,M^2)\right\},\\
 && \frac1{i}\int_I \frac{d^d k}{(2\pi)^d}\frac{ \{1,k^\mu, k^\mu k^\nu, k^\mu k^\nu k^\rho\} }
                          {(k^2-m_1^2+i\epsilon)(k^2-m_2^2+i\epsilon) \left[(p-k)^2-M^2+i\epsilon\right]}\nonumber\\
 &=& \left\{F_0(p^2,m_1^2,m_2^2,M^2),\quad
            p^\mu F_1(p^2,m_1^2,m_2^2,M^2),\quad
            g^{\mu\nu}F_2(p^2,m_1^2,m_2^2,M^2)+p^\mu p^\nu F_3(p^2,m_1^2,m_2^2,M^2),\quad
                 \right.\nonumber\\ &&\left.
            p^\mu p^\nu p^\rho F_4(p^2,m_1^2,m_2^2,M^2)
            +(g^{\mu\nu}p^\rho +g^{\mu\rho}p^\nu +g^{\nu\rho}p^\mu)F_5(p^2,m_1^2,m_2^2,M^2)\right\},
            \label{IRint}
\end{eqnarray}
where $M$ is the mass of the heavy meson and $m, m_1, m_2$ are the
masses of the light pseudoscalar mesons. The Lorentz invariant
coefficient $I^{(0)}$ can be written as \cite{Becher1999},
\begin{eqnarray}
  I^{(0)}(p^2,m^2,M^2)&=&-\frac{p^2-M^2+m^2}{p^2}L-\frac1{32\pi^2}\frac{p^2-M^2+m^2}{p^2}(2\log \frac{m}{\lambda}-1)
    \nonumber\\
  && +\frac{\alpha}{8\pi^2\tilde\omega^2}\times
       \begin{cases}
      -\sqrt{1-\Omega^2} \cos ^{-1}(-\frac{\Omega+\alpha}{\tilde\omega} )& |\Omega|\leq 1 \\
      \sqrt{\Omega^2-1} \log\left( \sqrt{(\frac{\Omega+\alpha}{\tilde\omega})^2-1}
                                  -\frac{\Omega+\alpha}{\tilde\omega}\right)  &  \Omega<-1\\
      \sqrt{\Omega^2-1} \left\{i\pi+\log\left( \frac{\Omega+\alpha}{\tilde\omega}
                              -\sqrt{(\frac{(\Omega+\alpha)^2}{\tilde\omega})^2-1} \right)\right\}&  \Omega> 1
     \end{cases},
\end{eqnarray}
where
\begin{equation}
  \Omega=\frac{p^2-m^2-M^2}{2M m}, \quad
  \alpha=\frac{m}{M}, \quad
  \tilde\omega=\sqrt{1+2\alpha \Omega+\alpha^2},\quad
  L=\frac{\lambda^{D-4}}{16\pi^2}\left\{\frac1{D-4}+\frac12(\gamma_E-1-\ln 4\pi)\right\}.
\end{equation}
The other coefficients are
\begin{eqnarray}
  I^{(1)}(p^2,m^2,M^2)&=&\frac{p^2-M^2+m^2}{2p^2}I^{(0)}(p^2,m^2,M^2)+\frac{m^2}{p^2}(L+\frac1{16\pi^2}\log\frac m\lambda)
                 ,\nonumber\\
  I^{(2)}(p^2,m^2,M^2)&=&\frac1{d-1}\left(m^2 I^{(0)}(p^2,m^2,M^2)-\frac{p^2-M^2+m^2}2 I^{(1)}(p^2,m^2,M^2)\right)
                 ,\nonumber\\
  I^{(3)}(p^2,m^2,M^2)&=&\frac{p^2-M^2+m^2}{2p^2} I^{(1)}(p^2,m^2,M^2)-\frac1{p^2}I^{(2)}(p^2,m^2,M^2),\nonumber\\
  F_j(p^2,m_1^2,m_2^2,M^2)&=&\frac{I^{(j)}(p^2,m_1^2,M^2)-I^{(j)}(p^2,m_2^2,M^2)}{m_1^2-m_2^2},\qquad j=0,1,2,3,
                  \nonumber\\
  F_4(p^2,m_1^2,m_2^2,M^2)&=&\frac1{p^2 d} \left( \frac{d+2}2 I^{(3)}(p^2,m_2^2,M^2)
                             + \frac{d+2}2 (p^2-M^2+m_1^2) F_3(p^2,m_1^2,m_2^2,M^2)\right.\nonumber\\ &&\left.
                             -2I^{(1)}(p^2,m_2^2,M^2)
                             -2m_1^2 F_1(p^2,m_1^2,m_2^2,M^2)\right)
                 ,\nonumber\\
  F_5(p^2,m_1^2,m_2^2,M^2)&=&\frac14 I^{(3)}(p^2,m_2^2,M^2) +\frac{p^2-M^2+m_1^2}4  F_3(p^2,m_1^2,m_2^2,M^2)
                              -\frac{p^2}2 F_4(p^2,m_1^2,m_2^2,M^2).
\end{eqnarray}

\par
With the definitions,
\begin{eqnarray}
  h_1(m)&\equiv & \frac{g^2\left\{2 M_D^2 F_5(M_{D^*}^2, m^2, m^2, M_{D}^2)  +
                        4 M_{D^*}^2 F_5(M_{D^*}^2, m^2, m^2, M_{D^*}^2) \right\} }{f^4} ,\nonumber\\
  h_2(m)&\equiv &\frac{g^2 M_D^2}{f^4 M_{D^*}}
               \partial_x I^{(2)}(x^2,m^2,M_D^2)|_{x\rightarrow M_{D^*}}
            +\frac{2g^2 M_{D^*}}{f^4}
               \partial_x I^{(2)}(x^2,m^2,M_{D^*}^2)|_{x\rightarrow M_{D^*}}   ,\nonumber\\
  h_3(m)&\equiv &\frac{g^2 M_D}{f^4 }
               \partial_x I^{(2)}(M_{D^*}^2,m^2,x^2)|_{x\rightarrow M_{D}}
            +\frac{2g^2 M_{D^*}}{f^4}
               \partial_x I^{(2)}(M_{D^*}^2,m^2,x^2)|_{x\rightarrow M_{D^*}}  ,
\end{eqnarray}
we can show the functions used for IR in Eq. (\ref{Tloop}),
\begin{eqnarray}
  W(m)&=&\frac{g^2}{f^4} \left\{4M_{D^*} F_2(M_{D^*},m,m,M_{D^*})+2M_D F_2(M_{D^*},m,m,M_{D})\right\}_r,  \nonumber\\
  J&=&\frac{g^2}{f^4} \left\{4M_{D^*} F_2(M_{D^*},m_\eta,m_\pi,M_{D^*})
               +2M_D F_2(M_{D^*},m_\eta,m_\pi,M_{D})\right\}_r,        \nonumber\\
  V(m,\omega)&=&\frac{3m^2 \omega}{8\pi^2 f^4}\log\frac m\lambda
        -\frac1{f^4}\left\{2M_{D^*} \omega^2  I^{(0)}(\omega+M_{D^*},m,M_{D^*})
        +4M_{D^*}^2\omega  I^{(1)}(\omega+M_{D^*},m,M_{D^*}) \right.\nonumber\\ && \left.
        \quad +2M_{D^*} I^{(2)}(\omega+M_{D^*},m,M_{D^*})
         +2M_{D^*}^3 I^{(3)}(\omega+M_{D^*},m,M_{D^*})\right\}_r  ,    \nonumber\\
  V_1&=&m_\pi\left\{-\frac12 h_1(m_K) -h_1(m_\pi) -\frac34 h_2(m_\pi) -\frac12 h_2(m_K) - \frac1{12} h_2(m_\eta)
         + \frac14 h_3(m_\pi) - \frac1{12} h_3(m_\eta) \right\}_r  ,\nonumber\\
  V_2&=&m_K\left\{-h_1(m_K)  - \frac23 h_2(m_K) -\frac29 h_2(m_\eta)
        + \frac13 h_3(m_K) - \frac29 h_3(m_\eta) \right\}_r  ,\nonumber\\
  V_3&=&m_K\left\{-h_1(m_\pi)  - \frac12 h_2(m_\pi) -\frac13 h_2(m_K) -\frac1{18} h_2(m_\eta)
        +\frac12 h_3(m_\pi)- \frac13 h_3(m_K) - \frac1{18} h_3(m_\eta) \right\}_r  ,\nonumber\\
  V_4&=&m_K\left\{  \frac32 h_2(m_\pi)- h_2(m_K) -\frac12 h_2(m_\eta)
        +\frac32 h_3(m_\pi)-  h_3(m_K) - \frac12 h_3(m_\eta) \right\}_r . \label{IRfunc}
\end{eqnarray}
The $\{X\}_r$ in Eq. (\ref{IRfunc}) represents $\displaystyle
\lim_{d\rightarrow 4} X$ after removing the terms proportional to
$L$.

\end{document}